\documentclass 
{aa}

\usepackage{aabib}

\newcommand{\f}{\frac}

\newcommand{\La}{\Lambda}

\newcommand{\ep}{\epsilon}

\begin{document}
\title{\bf Entropy \& equation of state (EOS) for hot bare strange stars}

\titlerunning{Entropy \& EOS for hot ... }

\author{ S. Ray $^{1,~ 2}$, J. Dey $^{2,~ 4~ \dagger}$,
M. Dey $^{1,~ 4 ~\dagger}$, K. Ray $^{1,~ 2}$  and B. C.
Samanta $^{3}$}
\authorrunning{Ray et~al.}

\institute{$^1$ Dept. of Physics, Presidency College, Calcutta 700
073, India\\ $^2$ Azad Physics Centre, Dept. of Physics, Maulana
Azad College, Calcutta 700 013, India\\ $^3$ Department of
Physics, Burdwan University, Burdwan 713104, India.\\ $^4$ Senior
Associate, IUCAA, Pune, India \\ $\dagger$ permanent address; 1/10
Prince Golam Md. Road, Calcutta 700 026, India;
e-mail:deyjm@giascl01.vsnl.net.in.
\\$^* $ Work supported   in  part  by DST  grant no. SP/S2/K18/96,
Govt. of India. }


\date{Received: \underline{28.09.2000}, Accepted:
\underline{22.11.2000}}

\thesaurus{02.04.1; 02.05.1; 02.05.2; 08.06.3}

\maketitle

\begin{abstract}
  {Compactness of some stars is explained if they are strange
stars (SS) as shown by Dey et  al. (1998) (D98) and Li et al.
(1999a). One of these compact star candidates is the
SAX~J1808.4$-$3658 (SAX in short) believed to be an important link
in the genesis of radio pulsars. SS  have also been suggested for
bursting X-ray pulsars (GRO~J1744$-$28, Cheng et al. 1998), from
quasi-periodic oscillations (QPO) of X-ray binaries (4U~1728$-$34,
Li et al. 1999b) and from peculiarity of properties of radio
pulsars (PSR~0943$+$10, Xu et al.1999; Kapoor et al. 2000). We now
extend the calculation to include high temperatures upto
$T$~=~70~MeV~$\sim~8\times 10^{11}~^o$K and find that the nature
of the mass (M) and radius (R), derived from astrophysical data,
is still retained. The entropy is calculated and matches onto that
calculated from hadronic models thus supporting the idea that the
quark-hadron transition may be continuous.}

\end{abstract}


\section{Introduction}


    A calculation for cold strange stars (D98) enabled us to draw
conclusions about chiral symmetry restoration in QCD when the EOS
was used to get SS fitting definite mass-radius (MR) relations
(D98; Li et al. 1999a; Li et al. 1999b; Ray et al. 2000b) . The
empirical MR relations were derived from astrophysical
observations like luminosity variation and some properties of
quasi-periodic oscillations from compact stars. The calculations
are compared to these stars which emit X-rays, generated
presumably due to accretion from their binary partner.

    During the genesis of these stars higher $T$ may be encountered and
in this paper we deal with a generalized case of an object at a
uniform $T$. We show at upto $T$ = 70 MeV a self sustained system
can be supported by the parameter set eos1 (D98) (called SS1 in Li
et al. 1999a).

    The changes in star masses are shown in the range of $T$ mentioned
above. The conclusions of D98 and Li et al. (1999a, 1999b) are
still valid for the finite $T$ cases. Secondly, parameters of the
single particle potential at finite $T$ are tabulated. In
particular the entropy is studied and this can be compared with
hadronic models. The comparison shows that the entropy may indeed
be continuous supporting the idea of a continuous phase transition
between hadrons and quarks.

\section{Strange stars at finite temperatures}

    The interesting point made in D98 is that starting
from an empirical form for the density dependent masses of the up
(u), down (d) and strange (s) q-s given below one can constrain
the parametric form of this mass from  recent astronomical
data{\footnote{ $\rho_{\rm B} = (\rho_{\rm u}+ \rho_{\rm d}+
\rho_{\rm s})/3$ is the baryon number density, $\rho_0 =
0.17~\rm{fm}^{-3}$ is the normal nuclear matter density, and $\nu$
is a numerical parameter. The current q masses $m_{\rm i}$ used in
the following are 4, 7 and 150 MeV for u, d and s respectively.}}:
\begin{equation}
M_{\rm i} = m_{\rm i} + M_{\rm Q}~~{\rm sech}(\nu \f{\rho_{\rm
B}}{\rho _0}), \;\;~~~ {\rm i = u, d, s}. \label{qm}
\end{equation}
Restoration of chiral quark masses at high density is incorporated
in this model. Using the model parameter ($\nu$) for this
restoration one can calculate the density dependence of the strong
coupling constant (Ray et al. 2000b). In other words the masses of
stars in units of solar mass, ($M/M_\odot$), found as a function
of the star radius $R$, calculated using the above Eq.~\ref{qm}, -
produces constraints which enable us to restrict the parameter
$\nu$. At high $\rho_{\rm B}$ the q- mass $M_{\rm i}$ falls from
its constituent value $M_{\rm Q}$ to its current one $m_{\rm i}$.
The parameter $M_{\rm Q}$ is taken to be 310 MeV to match up with
constituent q- masses assuming the known fact that the hadrons
have very little potential energy. The results are not very
sensitive in so far as changing $M_{\rm Q}$ to 320 MeV changes the
maximum mass of the star from 1.43735 $M_\odot$ to 1.43738
$M_\odot$ and the corresponding radius changes from 7.0553 kms to
7.0558 kms.

\input{psbox.tex}
\begin{figure}[htbp]
{\mbox{\psboxto(8cm;7cm){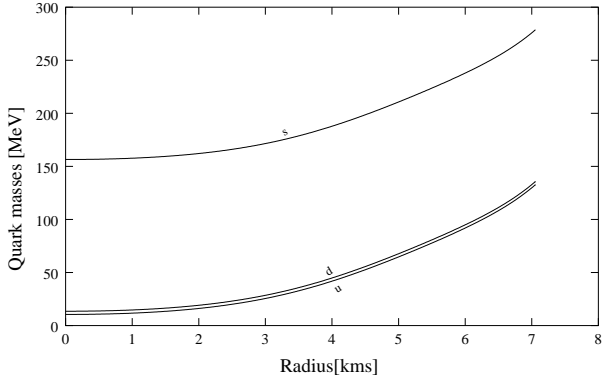}}} \caption{The smooth
restoration of chiral symmetry inside the star for each of the u,
d and s flavours (note that this is for the zero temperature
result).} \label{figchrl}
\end{figure}

    It is interesting to plot the up (u), down (d) and strange (s)
q- masses at various radii in a star. This is done with strong
coupling constant  $\alpha_0 = 0.2$, chiral symmetry restoration
parameter $\nu = 1/3$ and the QCD scale parameter $\La = 100$
already discussed in D98 \& Li et al. (1999a, 1999b).
Fig.~\ref{figchrl} shows that the quarks do not have the
constituent masses as in zero density hadrons nor do they have the
current masses of the bag model. Upto a radius about 2 kms the
quarks have their chiral current mass but in the major portion of
the star their masses are substantially higher. At the surface the
strange q- mass is about 278 MeV and the u, d q- masses $\sim 130$
MeV.

\begin{figure}[htbp]
{\mbox{\psboxto(8cm;7cm){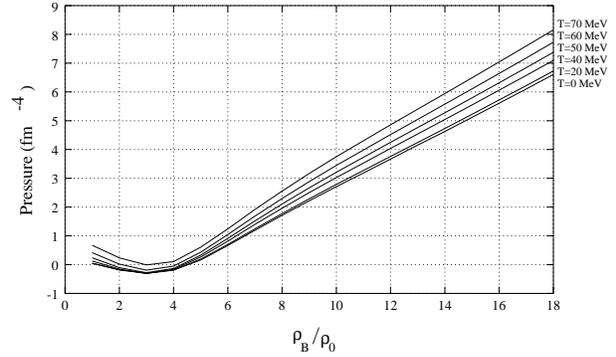}}} \caption{The equation
of states for different $T$. It is to be noted that beyond
$T$~=~70 MeV, the zero of the pressure cannot be attained.}
\label{eost}
\end{figure}


     We use the large $N_{\rm c}$ (colour) approximation of 't Hooft (1974)
for quarks, where quark loops are suppressed by $1/N_{\rm c}$ and
the calculation can be performed at the tree level with a mean
field derived from a qq interaction (Witten 1979). This was done
for baryons (Dey et al. 1986; Dey et al. 1991; Ray et al. 2000a)
and extended to dense systems like stars (D98). Following is the
Hamiltonian, with a two-body potential $V_{\rm ij}$ :
\begin {equation}
H = \sum_i ({\bf \alpha} _i . {\bf p}_i + \beta_i M_i) +
\sum_{i<j} \f{\lambda(i) .\lambda(j)}{4} V_{ij}. \label{eqH}
\end{equation}

    The vector potential in  Eq.~\ref{eqH} between quarks originate
from gluon exchanges, and the $\lambda$-s are the color SU(3)
matrices for the two interacting quarks. In the absence of an
accurate evaluation of the potential (e.g.~from large $N_{\rm c}$
planar diagrams) we borrow it from meson phenomenology,  namely
the Richardson potential (Richardson 1979). The potential
reproduces heavy as well as light meson spectra (Crater et al.
1984).  It has been well tested for baryons in Fock calculations
(Dey et al. 1986; Dey et al. 1991). Recently, using the Vlasov
approach with the Richardson potential, Bonasera (1999) finds a
transition from nuclear to quark matter at densities 5 times
$\rho_0$.

    We had to calculate the potential energy (PE) contribution in
two steps~: the single particle potential, $U_i(k)$, for momentum
$k$, is first calculated and this is subsequently integrated to
get the PE. The $U_i(k)$ is needed for doing finite $T$
calculation.

    The $U_i(k)$ is parametrized as sum of exponentials in $k$
(i.e., $U_i(k)= - {\rm exp} (a_0 + a_1 k + a_2 k^2)$ for a given
flavour), where the parameters for a given set, reported in this
paper, with $\nu = 1/3$, $\Lambda~ = ~100~~$MeV, $\alpha~ = ~0.2$
and a typical density $10 \rho_0$ are given in
Table~\ref{tabparam}.

\begin{table}[htbp]
\caption{Parameters of $U_i(k)= - {\rm exp} ({a_0}_i + {a_1}_i k_i
+ {a_2}_i k_i^2)$} \vskip 1cm
\begin{center}
\begin{tabular}{|c|c|c|c|}
\hline i & $a_0$ & $a_1$ & $a_2$ \\

\hline

u & $-$2.3524  & 0.10162 & $-$0.09832 \\

\hline

d & $-$2.3486 & 0.10281 & $-$0.10003 \\

\hline

s & $-$2.3109 & 0.12042& $-$0.11023\\

\hline
\end{tabular}
\end{center}
\label{tabparam}
\end{table}

        Finite temperature $T(=1/\beta)$ can be incorporated in the
system through the Fermi function~:
\begin{equation}
\rm FM({\it k,T})~=~\f{1}{\rm{exp}[{\it \beta (\ep - \ep_F)}] + 1}
\end{equation}
with the flavour dependent single particle energy
\begin{equation}
\ep_i~=~\sqrt{k^2+M_i(\rho)^2}+U_i(k). \label{eq:ep}
\end{equation}
Now we evaluate
 $$ \f{\gamma}{2\pi^2}\int_0^\infty
\phi(\ep)k^2\rm FM({\it k,T}) d{\it k}$$

For $\phi(\ep) = 1 $ we get the number density and for $\phi(\ep)
= \ep  $, the energy density. $\gamma = 6$ is the spin-colour
degeneracy.

    Even at very high $T$ which is around 70 MeV, the chemical
potential is very large, of the order several hundred MeV. The
entropy is calculated as
follows:$$s(T)~=~-\f{3}{\pi^2}\int_0^\infty {\it k}^2 \rm[FM({\it
k,T})\rm{ln}(FM({\it k,T}))~~~~~~~~~~~~~~~~~~~~$$
\begin{equation} ~~~~~~~~~~~~~~~~~+ \rm (1-FM({\it k,T}))ln(1-FM({\it k,T}))]
d{\it k} \label{eq:s}
\end{equation}
The pressure ($P$) is calculated from the free energy $$f = \ep -
Ts$$ as follows:
\begin{equation}
~~~~~~~~~P~=~\sum_i\rho_i\f{\partial f_i}{\partial \rho_i} - f_i
\label{pressure}
\end{equation}
We find $P$ = 0 points only upto $T$ = 70 MeV on plotting $P$ as a
function of $\rho$ at different $T$ (Fig.~\ref{eost}).

\section{Results and discussions}

\begin{figure}[htbp]
{\mbox{\psboxto(8cm;7cm){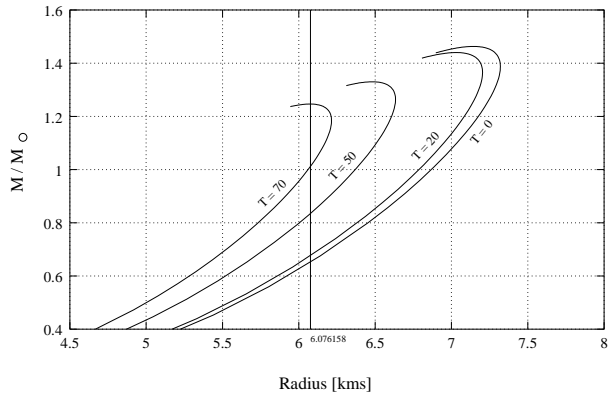}}} \caption{MR curves for
$T$~=~0, 20, 50 \& 70 MeV. Core densities for $T \ne 0$ are 16.5,
19.5 and 22 $\rho_0$. We have drawn a vertical line at $R_{\rm
max}$ for $T$~=~70 MeV to show that the mass of star increases as
$T$ increases, for fixed radius.} \label{mrft}
\end{figure}

    With this EOS at finite $T$, the TOV equation is solved to
find out the MR relationship of the strange star at finite $T$.
Here we find that the stellar mass and radius decrease with
increasing $T$ and the central energy density goes upto 22 times
the normal saturation density of nuclear matter. No star can be
formed beyond $T$~=~70 MeV since the star surface must have
pressure zero for SS. The MR curves for the stars at the three
values of $T$ are given in Fig.~\ref{mrft} showing that the radii
and the masses decrease slightly with $T$. The conclusions of D98
and Li et al. (1999a \& 1999b) remain unchanged so far as
compactness of the stars is concerned.

    We thank the anonymous referee for pointing out to
us that the decrease in the maximum mass and the corresponding
radius $R_{\rm max}$ of the star at finite $T$ is somewhat
surprising. We offer explanations of this fact as follows :

(1) We show in the Fig.~\ref{mrft} that the mass of the star with
a given radius increases with $T$. This is clear from the vertical
line which connects the maximum mass for $T$ = 70 MeV with $R_{\rm
max}~\equiv~6.076~$ km to the star mass for $T$ = 0 for the same
radius. This is in accord with the usual expectation that the star
mass should increase with $T$.

(2) With increasing $T$, the size of any self sustaining system
decreases due to the restriction placed on the energy balance by
the increased thermal energy. The shrinking of a self-sustained
system with increasing $T$ is also seen for the Skyrmion (Dey and
Eisenberg 1994). Interestingly the latter is also an example of
the success of the large $N_{\rm c}$ phenomenon. Note that $P$ in
our system (Fig.~\ref{eost}), or the Skyrmion, is calculated self
consistently whereas in the earlier literature on strange stars at
finite $T$ (Kettner et al. 1995) employing bag model, the
variation of the bag pressure with $T$ is neglected. It is pointed
out in (Chmaj and S{\l}omi\'nski 1989) at low $P$ the bag pressure
dominates and as $P$ grows the results are close to the free
relativistic gas limit. With an unchanged bag parameter therefore
bag model calculations find an almost unchanged mass and radius
for the stars at finite $T$.

(3) As thermal energy increases the binding energy per baryon
decreases. This supports the argument given in (1) and (2) above
that as $T$ decreases the masses corresponding to the same radius
will decrease.

    The entropy $S$ increases with $T$ and decreases with density.
Comparable to our entropy is that calculated by Das, Tripathi and
Cugnon (1986) (DTC) for interacting hadrons. The results of DTC
checks with experiment. The minimum $T$ considered by DTC is 10
MeV and $S/A$ is a little less than 1 at a density 2.5 times
$\rho_0$.

\begin{figure}[htbp]
{\mbox{\psboxto(8cm;7cm){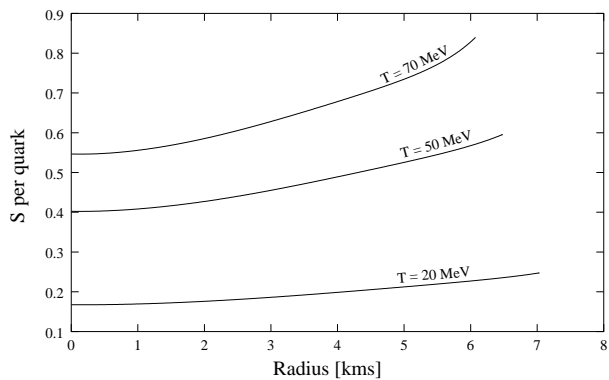}}} \caption{The entropy
per quark at different $T$(20, 50 \& 70 MeV) as a function of the
star radius.} \label{sradt}
\end{figure}

    In Fig.~\ref{sradt} we have plotted $S/A$ as a function of the
star radius for $T$~=~20, 50 and 70 MeV. It is interesting to see
here that entropy is maximum at the surface showing that the
surface is more disordered than the core. The extrapolated
entropies from the plots of DTC at 5 $\rho_0$ agree with our $S/A$
at the stellar surface for all the three cases.

Recently Glendenning (Glendenning 2000) has argued that the SAX
could be explained as a neutron star rather than bare SS, not with
any of the existing known EOS, but with one based only on
well-accepted principles and having a core density about 26
$\rho_0$. Of course, such high density cores imply hybrid strange
stars, subject to Glendenning's assumption that such stars can
exist with matching EOS for two phases. There is the further
constraint that if the most compact hybrid star has a given mass,
all lighter stars must be larger. It was found in Li et al.
(1999b) that the star 4U~1728$-$34 may have a mass less than that
of SAX and yet have a radius less than $R_{\rm SAX}$. Another
serious difference is that in our model the strange quark matter
EOS derived, using the formalism of large $N_{\rm c}$
approximation, indeed shows a bound state in the sense of having
minimum at about $4.8~\rho_0$ whereas in Glendenning (2000) one of
the assumptions is that strange matter has no bound state.

    In summary, we find that beyond $T$~=~70 MeV, the EOS has no
zero pressure point. A self-bound star cannot exist in our model
at higher $T$. The entropies at $T$~=~20, 50 and 70 MeV
(intermediate values are quite obvious) match onto hadronic
entropies at corresponding $T$, suggesting the possibility of
smooth phase transition between the hadronic and the quark states.

\acknowledgements The authors SR, JD, KR and MD are grateful to
IUCAA, Pune, India, for a short stay. It is also our great
pleasure to thank Prof. C. S. Shukre of the Raman Research
Institute, India, for important discussions.
\bibliographystyle{aabib}

\end{document}